\documentstyle[twoside,fleqn,espcrc2]{article}

\newcommand{\AmS}{{\protect\the\textfont2
  A\kern-.1667em\lower.5ex\hbox{M}\kern-.125emS}}

\title{Transport Coefficients of Quark Gluon Plasma 
       for Pure Gauge Models
       \thanks{Presented by S.Sakai}
}

\author{
     A.Nakamura,\address{Faculty of Education, 
        Yamagata University, Yamagata\\} 
     S.Sakai$^{\rm a}$ 
    and K.Amemiya\address{Research Center for Nuclear Physics,Osaka University,
         Ibaraki, Osaka}%
}

\begin{document}

\begin{abstract}
 The transport coefficients of quark gluon plasma are calculated 
on a lattice $16^3\times8$, with the pure gauge models.
Matsubara Green's functions of energy
momentum tensors have very large fluctuations and about a few 
million MC sweeps are needed to reduce the errors reasonably small in the 
case of the standard action. They are much suppressed if 
Iwasaki's improved action is employed.
Preliminary results show that the 
transport coefficients roughly depend on the coupling constant
as $a^{-3}(g)$.

\end{abstract}

\maketitle

\section{Introduction and Formulation}

Quark gluon plasma(QGP) is expected to be realized in high energy
heavy ion collisions in near future and thought to play an important
 roll
at the very early stage of the universe.
 Phenomenologically it will be
treated as a fluid and it will be very useful to  
calculate the transport coefficients from the
fundamental theory of QCD. 

  The calculation of the transport coefficients of QGP has been done
within the Kubo's linear response theory\cite{Hosoya,Karsch,Horsley}. 
According to this theory, the transport coefficients are calculated
by the space time integral of retarded Green's function at finite temperature. 
It is generally difficult to calculate the retarded Green's function
at finite temperature directly.
Then the shortcut is that the Matsubara Green's function is calculated 
and by analytic continuation retarded Green's function is obtaied. 
For the analytic
continuation the spectral representation of the both Green's functions is
used, for which the following ansatz is assumed\cite{Karsch},   
$$\rho(\vec{p}=0,\omega)=\frac{A}{\pi}                                       
(\frac{\gamma}{(m-\omega)^2+\gamma^2}-\frac{\gamma}{(m+\omega)^2+\gamma^2}),$$
\noindent
where $\gamma$ is related to the imaginary part of self energy.
From these parameters, the transport coefficients are calculated as,
$$ \alpha(A,\gamma,m) = 2A\frac{2\gamma m}{(\gamma^2+m^2)^2}, $$
here $\alpha$ represents heat conductivity,
bulk and shear viscosities.

We notice that if $m=0$ or $\gamma=0$, transport coefficient
becomes zero. 
In order to fix these parameters, at least three 
independent data points in temperature direction
are required for the Matsubara Green's function.

  In the pioneering work of Karsch and Wyld\cite{Karsch}, they performed
lattice QCD calculations on $8^3 \times 4$ lattice and the resolution was 
not enough to determine these three
parameters independently.

\section{Matsubara Green's Function}
We carry out the simulation on $16^{3} \times 8$ lattice for 
U(1),SU(2) and SU(3) pure gauge theories.
In the gauge theory, the energy momentum tensor is expressed as
follows,
$$T_{\mu\nu} = 2Tr(F_{\mu\sigma}F_{\nu\sigma}                        
-\frac{1}{4}\delta_{\mu\nu}F_{\rho\sigma}F_{\rho\sigma}),$$
\noindent
where $F_{\mu\nu}$ are field strength tensor defined by plaquette 
variables as 
$ U_{x,\mu\nu} = \exp{(ia^2gF_{\mu\nu})} $.            
From the plaquette variable $F_{\mu\nu}$ is obtained either by taking $log$ 
or by expanding with respect to $ga^2$. We call the former method as 
'diagonal method' and a latter one as 'perturbative method'.

   We have started the simulation from compact U(1) case, because it has
two phase, the confinement and the Coulomb phase which are separateded at 
$\beta \sim 1.0$, and is a good place to test the formulation. 
 In the following we denote the Matsubara Green's
function of energy momentum tensor $T_{\mu\nu}$ as $G_{\mu \nu}(T)$,

In Fig. 1, we show how the error decreases with number of MC sweeps in
confined and deconfined phase for U(1) case. In this work the mesurements
are done with every sweep. 
The difference in the fluctuations of
$G_{\mu \nu}(T)$ between the confined and deconfined phase are observed.
In the confined phase, we could not reduce the ratio
$Errors/<G(T)>$ less than unity within more than million data even at 
$T=2$.
Similar situation is observed in SU(2) case.
\begin{figure}[htb]
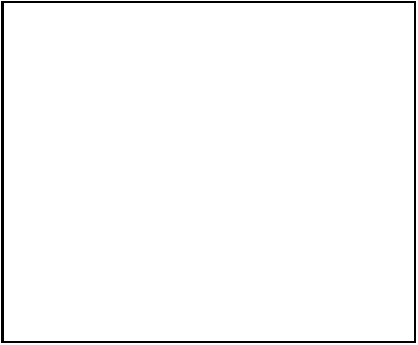

\vspace{9pt}
\framebox[55mm]{\rule[-21mm]{0mm}{43mm}}
\caption{Error as a function of number of MC sweeps.}
\label{fig:largenenough}
\end{figure}
We interpret this as follows; the energy momentum tensor in the 
confined phase should be written by the hadron fields.  Namely $T_{\mu \nu}$
used in this calculations are not good operators in the confined phase.
Therefore in the following we
calculate the transport coefficients only in the deconfined phase.

  The fluctuations of $G_{\mu \nu}(T)$ for U(1),SU(2) and SU(3) 
cases in 
the deconfined phase are shown in Fig.2.,
where transition point
is $ \beta \sim 6.05$ for SU(3)\cite{qcdt} and
cross over
region is $\beta \sim 2.4$ for SU(2) for $N_t=8$.  
The fluctuation becomes larger as we proceed to SU(2) 
and SU(3).
And as the fluctuation increases as $T$ becomes large, we need about
million data for SU(2) $\beta = 3.0$ case to make the ration 
$Error(T)/<G(T)> \leq 1$ for $T=4$, 
and more data will be needed for SU(3) case. 
  
  In order to decrease the fluctuation,
we try to use an improved action proposed
by Iwasaki\cite{Iwasaki} for the SU(3) case. 
The phase transition region
for the Iwasaki's improved action on $16^3 \times 8$ lattice 
is $ \beta \sim 2.7-2.9$ and we started our simulation at
$\beta=3.3$ .
In Fig.2 we have also shown how errors decrease with number of 
MC sweeps for the improved action. 
The fluctuation is much reduced by the use of 
the improved action. 
From Fig.2 we expect that within a few hundred thousand data, 
we could make
the ratio $Error(T)/<G(T)> \leq 1$ and get the  
transport coefficients of QGP with reasonable accuracy.
\begin{figure}[htb]
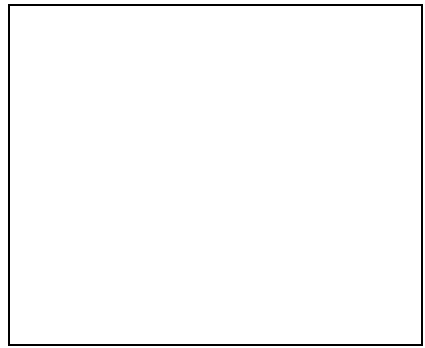

\framebox[55mm]{\rule[-21mm]{0mm}{43mm}}
\caption{Error as a function of number of MC sweeps at $T=2$ for 
U(1) $\beta=1.2$, SU(2) $\beta=3.0$, SU(3) $\beta=6.25$ and 
improved action for SU(3) $\beta=3.3$.}
\label{fig:toosmall}
\end{figure}

\section{Transport Coefficients}

 We determine the parameters in the spectral function by making the
fit of Matsubara Green's function. The fit of $G_{11}(T)$ is shown 
in Fig.3 for SU(2) at $\beta=3.0$, where  
the parameters and their errors are determined by the jackknife 
method using the least square package SALS.
In this work we excludes the points where $Error/<G(T)>$ larger than unity 
from the fit range, and the data at $T=0$ is not included.

    We could not find the difference between 
the definitions of field strength
tensor from the plaquette variables('diagonal' and 'perturbative' definitions)
in the present accuracy of data. Then at this $\beta$ region,
the higher order effects in $ga^{2}$ may be small for SU(2) gauge theory.
\begin{figure}[htb]
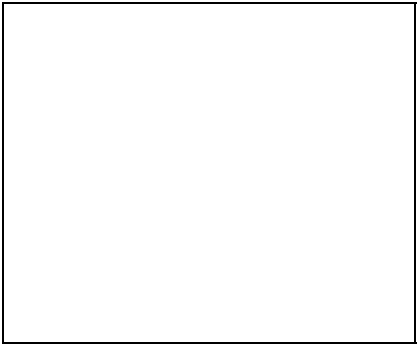

\vspace{9pt}
\framebox[55mm]{\rule[-21mm]{0mm}{43mm}}
\caption{Fit of $G_{11}(T)$ by the parameters of spectral function for
 SU(2) at $\beta=3.0$}
\label{fig:largenenough}
\end{figure} 

   In the regin $T = 3-4$, $G_{11}(T)$ becomes 
very flat, which is also observed for U(1) and SU(3).
This behavior could not be found if we do the
simulation in the smaller lattice of $N_{t}=4$. \\
\indent
The sheer viscosity $\eta$ and bulk viscosity $\xi$
are separated by taking the following combination,
 $$\eta = \alpha(A,\gamma,m)_{12},$$
 $$\xi = \alpha(A,\gamma,m)_{12} -\frac{4}{3}\alpha(A,\gamma,m)_{11},$$
where the subscript means that the transport coefficients determined
from $G_{\mu \nu}$.  
The sheer and bulk viscosities are shown in Fig.4 for SU(2) case.
The errors are estimated by 
jackknife method.

\begin{figure}[htb]
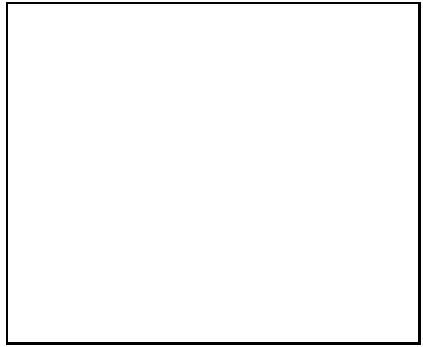

\vspace{9pt}
\framebox[55mm]{\rule[-21mm]{0mm}{43mm}}
\caption{The sheer and bulk viscosity of SU(2) case as a function of $\beta$}
\label{fig:largenenough}
\end{figure} 

  The calculation for the SU(3) case is now under simulation at 
$\beta=3.3$ and $3.2$. The result for the transport coefficients 
at $\beta=3.3$ are
$\eta \times a^3 =0.93 \times 10^{-3} \pm 0.107\times10^{-2}$ , and 
$\xi \times a^3 = 0.10\times10^{-2} \pm 0.13\times10^{-2}$. 
The results have still 
large errors and are very preliminary, but
we think that within a few month the result will be improved and the data
at $\beta=3.2$ will also be presented.\\  
\indent                   
  We have found that, in the confined phase, it is very difficult to
reduce the error of $G_{\mu\nu}$. We interpret it because the energy momentum 
tensor in the 
confined phase should be written by the hadron fields.  
In the deconfined phase the fluctuation for $G_{\mu\nu}$ is still large,
but by the use of Iwasaki's improved action, it is much suppressed and 
the transport coefficients for SU(3) gauge
theory is calculated on $16^3 \times 8$ lattice.
 However there are many thing to do before we obtain the result for
the transport coefficients which
are used for the quantitative phenomenology of quark gluon plasma.

\bigskip
\noindent ACKNOWLEDGEMENT  
Calculations of the SU(2) and SU(3) part have been done on 
VPP/500 at KEK and by AP1000 at Institute for Nuclear
Study. We would like to express our                 
thanks to the members of KEK and INS for their warm hospitality.

\end{document}